\documentclass[aps,prl,color,pdflatex,citeautoscript,superscriptaddress,twocolumn]{revtex4-1}
\usepackage[T1]{fontenc}
\usepackage{graphicx}
\usepackage{dcolumn}
\usepackage{comment}
\usepackage{amsmath,amssymb,bbold,bm}
\usepackage{hyperref}
\hypersetup{colorlinks=true,citecolor=blue,linkcolor=blue,urlcolor=blue,pdfstartview=FitH,pdfpagemode=PageWidth}
\usepackage{amsfonts}
\usepackage{listings}
\usepackage{braket}
\usepackage{float,latexsym}
\usepackage{multirow}
\usepackage{tikz}
\usepackage{color}

\begin{document}
\title{Time-crystalline topological superconductors}

\author{Aaron Chew}
\affiliation{Department of Physics and Institute for Quantum Information and Matter, California Institute of Technology, Pasadena, CA 91125, USA}
\author{David F. Mross}
\affiliation{Department of Condensed Matter Physics, Weizmann Institute of Science, Rehovot, 76100, Israel}
\author{Jason Alicea}
\affiliation{Department of Physics and Institute for Quantum Information and Matter, California Institute of Technology, Pasadena, CA 91125, USA}
\affiliation{Walter Burke Institute for Theoretical Physics, California Institute of Technology, Pasadena, CA 91125, USA}

\date{\today}

\begin{abstract}

Time crystals form when arbitrary physical states of a periodically driven system spontaneously break discrete time-translation symmetry. 
We introduce one-dimensional time-crystalline topological superconductors, for which time-translation symmetry breaking and topological physics intertwine---yielding anomalous Floquet Majorana modes that are not possible in free-fermion systems.  
Such a phase exhibits a bulk magnetization that returns to its original form after two drive periods, together with Majorana end modes that recover their initial form only after four drive periods. 
We propose experimental implementations and detection schemes for this new state.
\end{abstract}

\maketitle

{\bf \emph{Introduction.}}~Periodically driven quantum systems evade certain constraints faced by their equilibrium counterparts.  
For instance, `time crystals' that spontaneously break time-translation symmetry in the sense envisioned in Refs.~\onlinecite{WilczekCrystal, ShapereCrystal} cannot arise in equilibrium \cite{WatanabeNoGo}, yet can emerge with periodic driving.  
Periodically driven time crystals exhibit the striking property that \emph{any} physical (i.e., non-cat) state evolves with a subharmonic of the drive frequency \cite{KhemaniGlass, ElseCrystal, YaoCrystal}.  
The canonical realization consists of disordered Ising spins that collectively flip after each drive period, thereby requiring two periods to recover their initial state.  
Experiments have detected signatures of time crystallinity both in driven cold atoms \cite{ZhangExp,SmitsExp} and solid-state spin systems \cite{ChoiExp, PalExp, RovnyExp}.

As a second, deeply related example, consider a one-dimensional (1D) free-fermion topological superconductor hosting Majorana end modes \cite{KitaevWire}, each described by a Hermitian operator $\gamma$.   If $\gamma$ adds energy $E$ then $\gamma^\dagger$ adds $-E$, while Hermiticity requires that these be equivalent.
In equilibrium the unique solution is $E = 0$---corresponding to the well-studied Majorana zero modes.
Periodically driving with frequency $\Omega$ additionally permits `Floquet Majorana modes' carrying $E = \Omega/2$ since energy is then only conserved mod $\Omega$ \cite{JiangMajorana}.  
Floquet Majorana modes have been proposed to facilitate more efficient quantum information processing compared to equilibrium systems \cite{BomantaraQC, BomantaraQC2, BauerQC}.  
Moreover, they encode a topological flavor of time-translation symmetry breaking in that Floquet Majorana operators change sign each drive cycle, thus also requiring two periods to recover their initial form.

\begin{figure}
\includegraphics[trim=1.9cm 0cm 0cm 0cm,clip=true,scale=0.30]{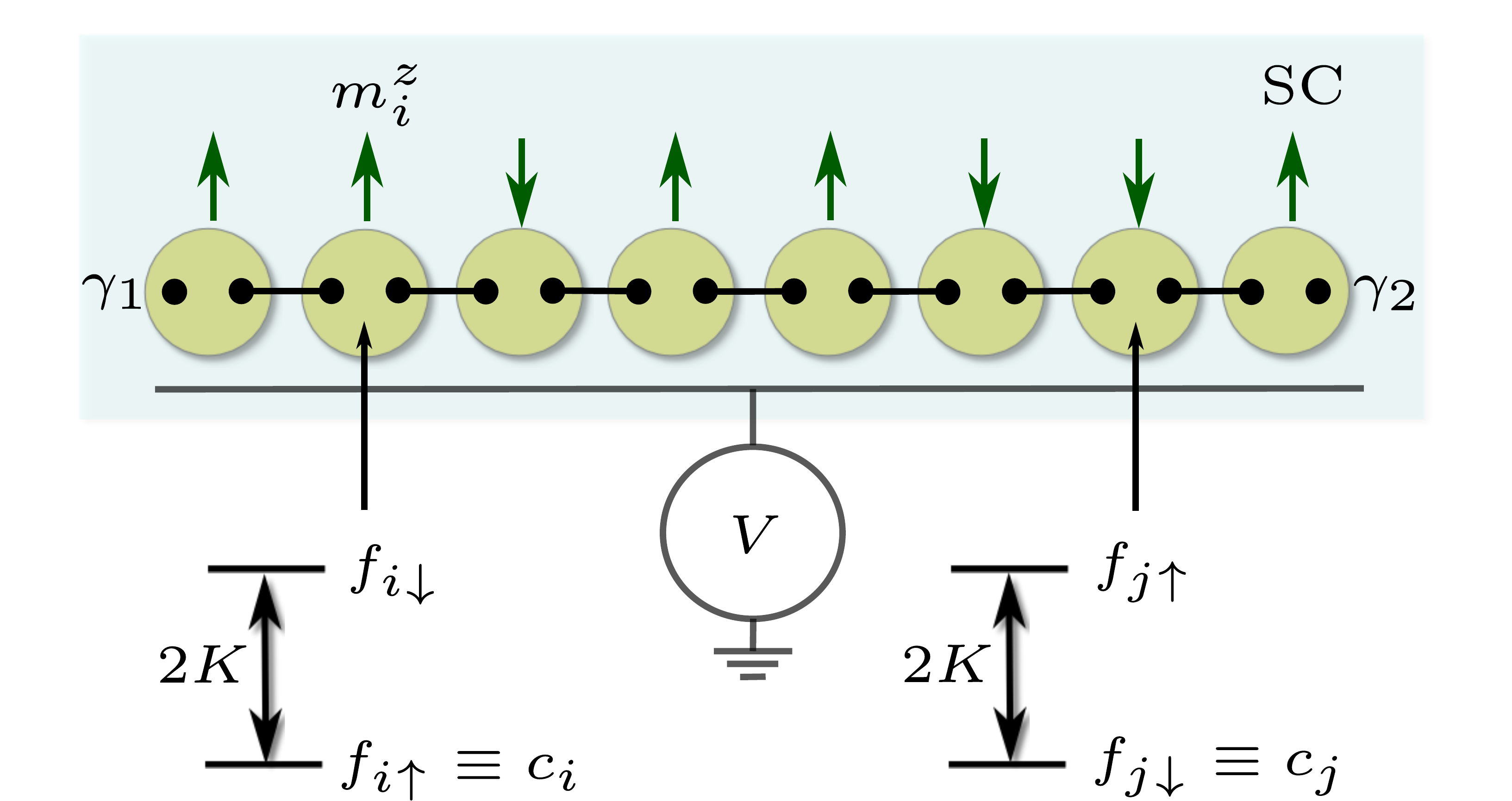}
\caption{Proximitized quantum-dot array coupled to Ising spins.  The Ising spins polarize the dot electrons---effectively producing a system of spinless fermions $c_j$.  In any Ising configuration, the fermions can realize topological superconductivity with unpaired Majorana zero modes $\gamma_{1,2}$ that intertwine with the adjacent spins.  }
\label{SetupFig}
\end{figure}

We merge the phenomena above by exploring periodically driven 1D topological superconductors generated upon coupling Cooper-paired electrons to doubled-periodicity time-crystalline Ising spins. 
Such `time-crystalline topological superconductors' intertwine bulk time-translation symmetry breaking and topological physics---yielding anomalous \emph{quadrupled-periodicity} Floquet Majorana modes that categorically can not arise in free-fermion platforms.
As a concrete implementation, we focus on quantum-dot arrays (see Fig.~\ref{SetupFig}) reminiscent of setups utilized in Refs.~\onlinecite{ChoyDot, SauDot, FulgaDot} for engineering equilibrium Majorana zero modes.  
We derive and analyze an exactly solvable, physically intuitive model for time-crystalline topological superconductivity 
and show that the quadrupled periodicity of the Floquet Majorana modes can be experimentally revealed by probing junctions between time-crystalline and static topological superconductors. 

{\bf \emph{Model and Setup.}}~The nontrivial properties of time-crystalline topological superconductors closely relate to equilibrium physics of topological superconductors that spontaneously violate electronic time-reversal symmetry $\mathcal{T}$, which importantly satisfies $\mathcal{T}^2 = -1$.  
We thus begin by exploring a time-independent model for the latter. 
Our setup, sketched in Fig.~\ref{SetupFig}, consists of a superconductor coupled to a chain of quantum dots indexed by sites $j$, each hosting one active spinful level described by operators $f_{j\sigma}$ ($\sigma = \uparrow,\downarrow$ denotes spin); we assume that charging energy is quenched by coupling to the superconductor and can thus be neglected.  
A chain of Ising spins described by Pauli matrices $m^z_j$ resides proximate to the quantum-dot array.   
We model the setup with a $\mathcal{T}$-symmetric Hamiltonian $H = H_0 + H_f$, where
\begin{align}
H_0 &= \sum_j ( -J m^z_jm^z_{j+1} - K m^z_j  f_j^\dagger \sigma^z f_j ), \label{Hline1} \\
H_f &= \sum_j [ -\mu f_j^\dagger f_j  -t (f_j^\dagger f_{j+1} + H.c.) \nonumber \\ 
&+ \alpha(if_j^\dagger \sigma^x f_{j+1} + H.c.)  + \Delta (f_{j\uparrow} f_{j\downarrow} + H.c.) ].
\label{H}
\end{align}
In $H_0$,  $J>0$ ferromagnetically couples neighboring Ising spins and $K>0$ couples the Ising and dot spins.
Terms in $H_f$ describe the chemical potential ($\mu$), hopping ($t$), spin-orbit coupling ($\alpha$), and proximity-induced pairing ($\Delta$) for the quantum-dot electrons.

Suppose that the $K$ term dominates and energetically enforces alignment of each electron spin with the nearest Ising spin.  
Only one of the two spinful levels in each dot remains active at low energies---effectively creating a system of spinless fermions described by operators
\begin{equation}
  c_j = \frac{1}{2}\left[(1+m_j^z)f_{j\uparrow} + (1-m_j^z)f_{j\downarrow}\right],
  \label{cj}
\end{equation}
as Fig.~\ref{SetupFig} illustrates.
Time-reversal ${\mathcal T}$ sends $m_j^z \rightarrow -m_j^z$ and $c_j \rightarrow m_j^z c_j$, thus conforming to $\mathcal{T}^2 = -1$ in the fermionic sector.  The intertwinement between spinless fermions and Ising spins evident here is unavoidable; without it, $c_j$ has no way of acquiring the required minus sign upon two applications of ${\cal T}$.

Appendix~\ref{ProjectionAppendix} projects $H$ onto the spinless-fermion subspace by integrating out high-energy fermionic modes,  
yielding an effective Hamiltonian
\begin{align}
H_{\rm eff} &= \sum_j [-J m_j^zm_{j+1}^z -\mu' c_j^\dagger c_j 
\nonumber \\
 &+ (t'_{m_j^z,m_{j+1}^z} c_j^\dagger c_{j+1}  + \Delta'_{m_j^z,m_{j+1}^z} c_j c_{j+1} +H.c.)].
\label{StaticHamiltonian}
\end{align} 
Here $\mu' = -(K+\mu)$ is a renormalized chemical potential, while $t'_{m_j^z,m_{j+1}^z} = a + a^* m_j^z m_{j+1}^z$ and $\Delta'_{m_j^z,m_{j+1}^z} = b m_j^z - b^* m_{j+1}^z$ denote Ising-spin-dependent effective hopping and $p$-wave pairing amplitudes, with $a = (-t+i \alpha)/2$ and $b = (-t+i\alpha) \Delta/(K-\mu)$. 
The real part of $a$ sets the hopping strength between sites with aligned Ising spins, which is directly inherited from spin-conserving tunneling in Eq.~\eqref{H}; the imaginary part similarly fixes the hopping when Ising spins anti-align, which is instead mediated by spin-orbit coupling $\alpha$.  
Pairing in $H_{\rm eff}$ follows from second-order processes that involve virtual excitations out of the spinless-fermion subspace---hence the $K-\mu$ energy denominator in $b$.  
Depending on the Ising configuration, either spin-conserving hopping or spin-orbit coupling virtually creates a doubly-occupied site of $f$ fermions that then Cooper pair via the original $s$-wave $\Delta$ term, effectively mediating $p$-wave pairing of spinless fermions. 

\begin{figure}
\includegraphics[width = .49\columnwidth]{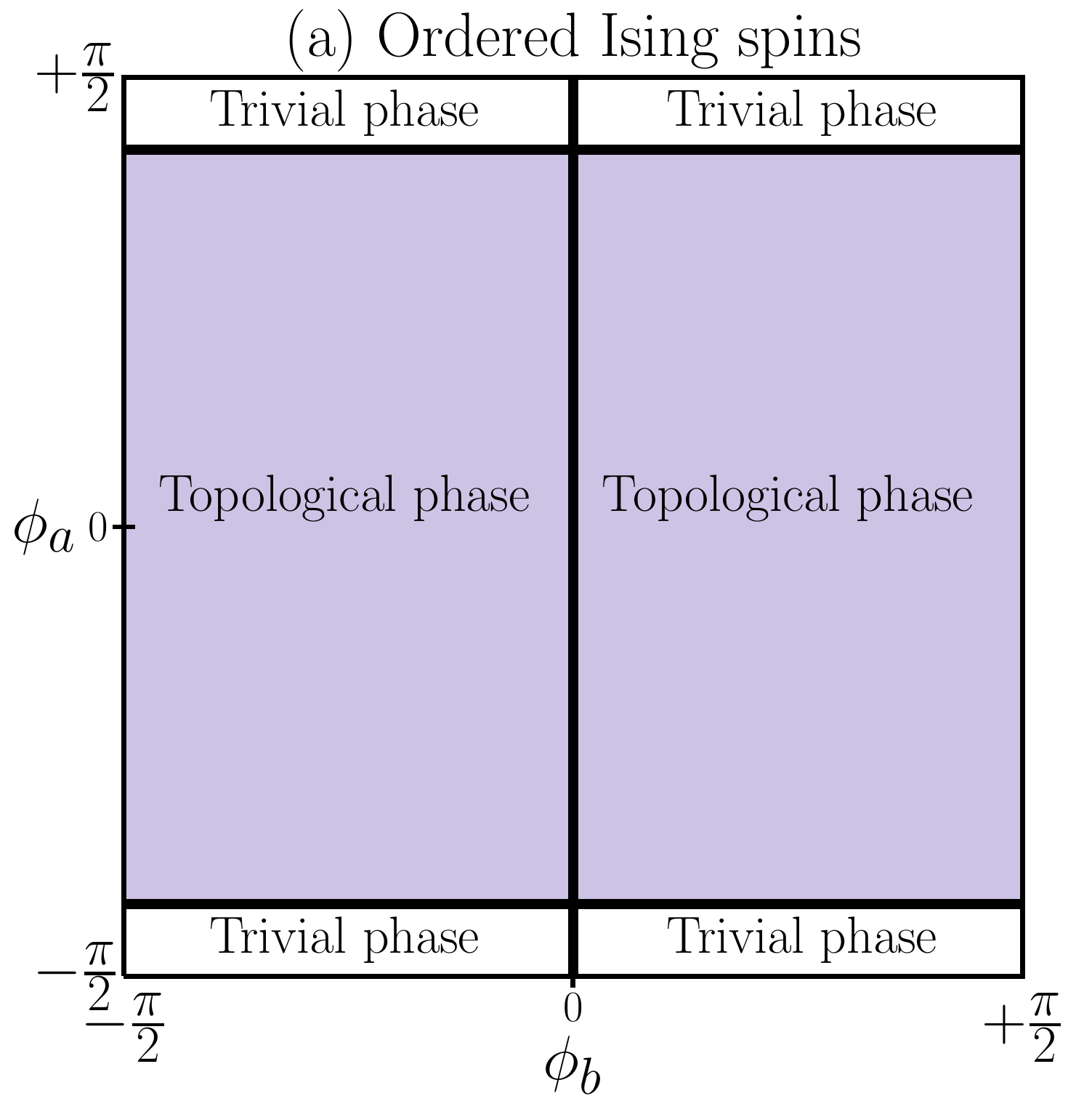}
\includegraphics[width = .49\columnwidth]{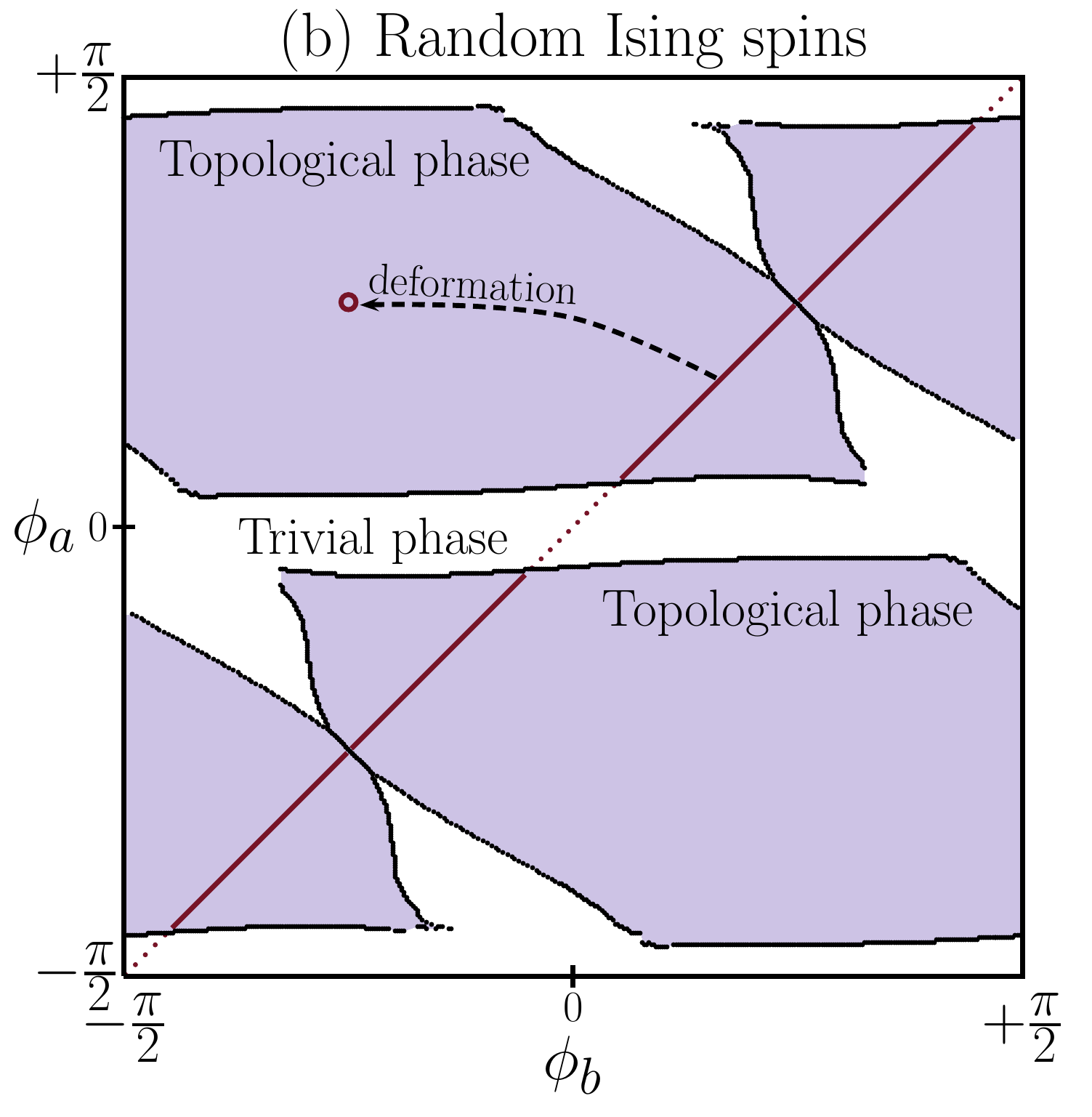}
\caption{Phase diagram for Eq.~\eqref{StaticHamiltonian} assuming (a) fully polarized and (b) random Ising spins.  
In (a) a nonzero chemical potential $\mu'=|a|$ generates the trivial phase, and the system is gapless along the thick black lines.
Data in (b) were generated from transfer-matrix simulations at $\mu' =| b| = |a|/4$ with $10^6$ sites. Data points indicate sharp peaks in the localization length, as expected at a topological phase transition. The red diagonal line $\phi_a = \phi_b$ is relevant for the physical quantum-dot setup from Fig.~\ref{SetupFig}.
As the dashed arrow illustrates, the topological phase along this line can be deformed to the zero-correlation-length limit with $\phi_a = \pi/4, \phi_b = -\pi/4$ (and also $|a| = |b|$, $\mu' = 0$) without crossing a phase boundary.
}
\label{FreeFermionFig}
\end{figure}

{\bf \emph{Phase Diagram.}}~Equation~\eqref{StaticHamiltonian} describes a strongly interacting system of Ising spins and fermions.
Nevertheless, for any given Ising configuration the model reduces to free fermions.
Any Ising configuration also breaks time-reversal symmetry, thus allowing the fermions to realize topological superconductivity with unpaired Majorana zero modes.  
Consider first uniformly polarized all-up or all-down Ising spins.  
Here Eq.~\eqref{StaticHamiltonian} maps to the familiar Kitaev chain \cite{KitaevWire} with uniform hopping strength $2|a| \cos\phi_a$ and pairing $\pm 2i|b|\sin{\phi_b}$, where $a = |a|e^{i\phi_a}$ and $b = |b|e^{i\phi_b}$.  
(Our derivation above yielded $\phi_a = \phi_b$, though it will be useful to now keep these phases independent.)
Accordingly, the chain hosts edge Majorana zero modes provided the chemical potential intersects the band and pairing is finite, i.e., for $|\mu'| < 4 |a| |\cos\phi_a|$ and $\sin\phi_b \neq 0$ as sketched in Fig.~\ref{FreeFermionFig}(a).  

To examine the fermionic ground state with random Ising spins---which is our main interest---we compute the correlation length $\xi$ using the transfer-matrix technique; see, e.g, Ref.~\onlinecite{MacKinnon2003} and Appendix~\ref{TransferMatrixApp}.  
This method allows us to map out phase boundaries by numerically searching for diverging $\xi$ as we vary $\phi_{a,b}$; for our purposes a regular 400$\times$400 grid of $\phi_a$ and $\phi_b$ values in the interval $[-\pi/2,\pi/2]$ is sufficient. [The number of simulations can be halved by virtue of the symmetry $\xi(\phi_a,\phi_b) = \xi(-\phi_a,-\phi_b)$]. Figure~\ref{FreeFermionFig}(b) illustrates representative results obtained for $\mu' = |b| = |a|/4$  and $N = 10^6$ sites. The data points indicate local maxima where $\xi$ is typically of order $10^2$ or larger, while it is of order unity elsewhere. We expect these peaks to represent true divergences in $\xi$ when $\phi_a$ or $\phi_b$ are tuned continuously in the thermodynamic limit. Topological regions labeled in the figure are easily identified by performing exact diagonalization on smaller systems and confirming the presence of edge Majorana zero modes.
In Appendix~\ref{BornAppendix} we analytically capture the topological phase for a restricted window of $\phi_{a,b}$ via the Born approximation.

For our quantum-dot setup, we expect $\phi_a = \phi_b$ [red line in Fig.~\ref{FreeFermionFig}(b)] and also $|a| \gg |b|$ since $p$-wave pairing encoded in $b$ appears at second order in perturbation theory.  
Starting from the topological phase in this physical regime, Fig.~\ref{FreeFermionFig}(b) strongly suggests that we can deform parameters to $\phi_a = \pi/4$ and $\phi_b = -\pi/4$, $|a| = |b|$, and $\mu' = 0$ without encountering a divergent $\xi$.
(See Appendix~\ref{TransferMatrixApp} for additional evidence.)
This special point corresponds to the model's zero-correlation-length limit. 
Here it is convenient to decompose the spinless fermions in terms of Majorana operators $\eta_{A,Bj}$ via $c_j = e^{-i\frac{\pi}{4}m^z_j}(\eta_{Bj} + i\eta_{Aj})$, whereupon Eq.~\eqref{StaticHamiltonian} becomes
\begin{align}
H'_\text{eff} = \sum_j(-J m^z_j m^z_{j+1} 
-i\kappa s_{m^z_j,m^z_{j+1}} \eta_{Aj} \eta_{Bj+1})
\label{EffHamiltonian}
\end{align}
with $s_{m_i,m_j} = (1-m_i+m_j+m_i m_j)/2 = \pm 1$ and $\kappa = 4\sqrt{2} |a|$. 
\emph{For any choice of $m^z_j$'s} the Majorana operators dimerize nontrivially as shown in Fig.~\ref{SetupFig}, yielding  
Majorana zero modes 
\begin{eqnarray}
  \gamma_1 &\equiv& \eta_{B1} = e^{i\frac{\pi}{4}m^z_1} c_1 + H.c.
  \nonumber ~\\
  \gamma_2 &\equiv& \eta_{AN} = -ie^{i\frac{\pi}{4} m^z_N} c_N + H.c. 
  \label{gamma_def}~
\end{eqnarray}
at the leftmost and rightmost sites.
Notice the spin-fermion intertwinement inherent in the zero modes, which consequently evolve under $\mathcal{T}$ via
\begin{equation}
  \gamma_1 \rightarrow m^z_1 \gamma_1,~~~~ \gamma_2 \rightarrow -m^z_N \gamma_2,
  \label{TransformationRules}
\end{equation}
again consistent with $\mathcal{T}^2 = -1$.  
All Hamiltonian eigenstates are at least fourfold degenerate in this limit: one factor of two arises because $\mathcal{T}$ flips all Ising spins, while the other reflects topological degeneracy encoded in the Majorana zero modes.  
The topological degeneracy of the fermionic ground states given a static Ising configuration persists even away from the special limit examined above, due to the finite gap for fermionic excitations. 
Moreover, Appendix~\ref{SolutionAppendix} shows that Eq.~\eqref{TransformationRules} holds even when the zero-mode wavefunctions extend over many sites.

{\bf \emph{Adiabatic cycle.}}~Next we generalize Eq.~\eqref{Hline1} to 
\begin{equation}
  H_0' = \sum_j [-J ({\bf \hat{n}}\cdot {\bf m}_j)({\bf \hat{n}}\cdot {\bf m}_{j+1}) - K ({\bf \hat{n}}\cdot {\bf m}_j)  f_j^\dagger {\bf \hat{n}}\cdot\boldsymbol{\sigma} f_j ],
\end{equation}
where ${\bf m}, \boldsymbol{\sigma}$ denote vectors of Pauli matrices and the unit vector ${\bf \hat{n}} \equiv \cos\theta {\bf \hat{z}} + \sin\theta {\bf \hat{y}}$ determines the easy axis for the Ising spins.  
At either $\theta = 0$ or $\pi$, $H_0'$ reduces to Eq.~\eqref{Hline1}.  
Suppose that we again deform to the zero-correlation-length limit (which is possible for any $\theta$) and then implement the following cycle: $(i)$ Start with an arbitrary Ising spin configuration at $\theta = 0$, $(ii)$ initialize the fermions into one of the topological-superconductor ground states, and finally $(iii)$ adiabatically rotate the easy axis by winding $\theta$ from $0$ to $\pi$. 

Although the Hamiltonian returns to its original form, \emph{the wavefunctions do not}.  
Rather, the cycle slowly rotates all Ising spins by $\pi$, while the fermions follow their instantaneous minimum-energy configuration given the adiabaticity. 
The initial ground state thereby transforms into its time-reversed counterpart.  
More formally, the easy-axis rotation sends $m^z_j \rightarrow -m^z_j$, $f_j \rightarrow e^{i \frac{\pi}{2}\sigma^x}f_j$, and hence $c_j \rightarrow i c_j$.  
It follows that the Majorana zero modes transform as $\gamma_1\rightarrow m^z_1 \gamma_1$ and $\gamma_2 \rightarrow m^z_N \gamma_2$, similar to the action of $\mathcal{T}$. 
Interestingly, two cycles return the Ising spins to their original form whereas \emph{four cycles} are required to recover the initial zero-mode operators, e.g.,
\begin{equation}
  \gamma_1 \rightarrow m^z_1 \gamma_1 \rightarrow -\gamma_1 \rightarrow -m^z_1\gamma_1 \rightarrow \gamma_1.
  \label{FourAdiabaticCycles}
\end{equation}  

\begin{figure}
\includegraphics[width = \columnwidth]{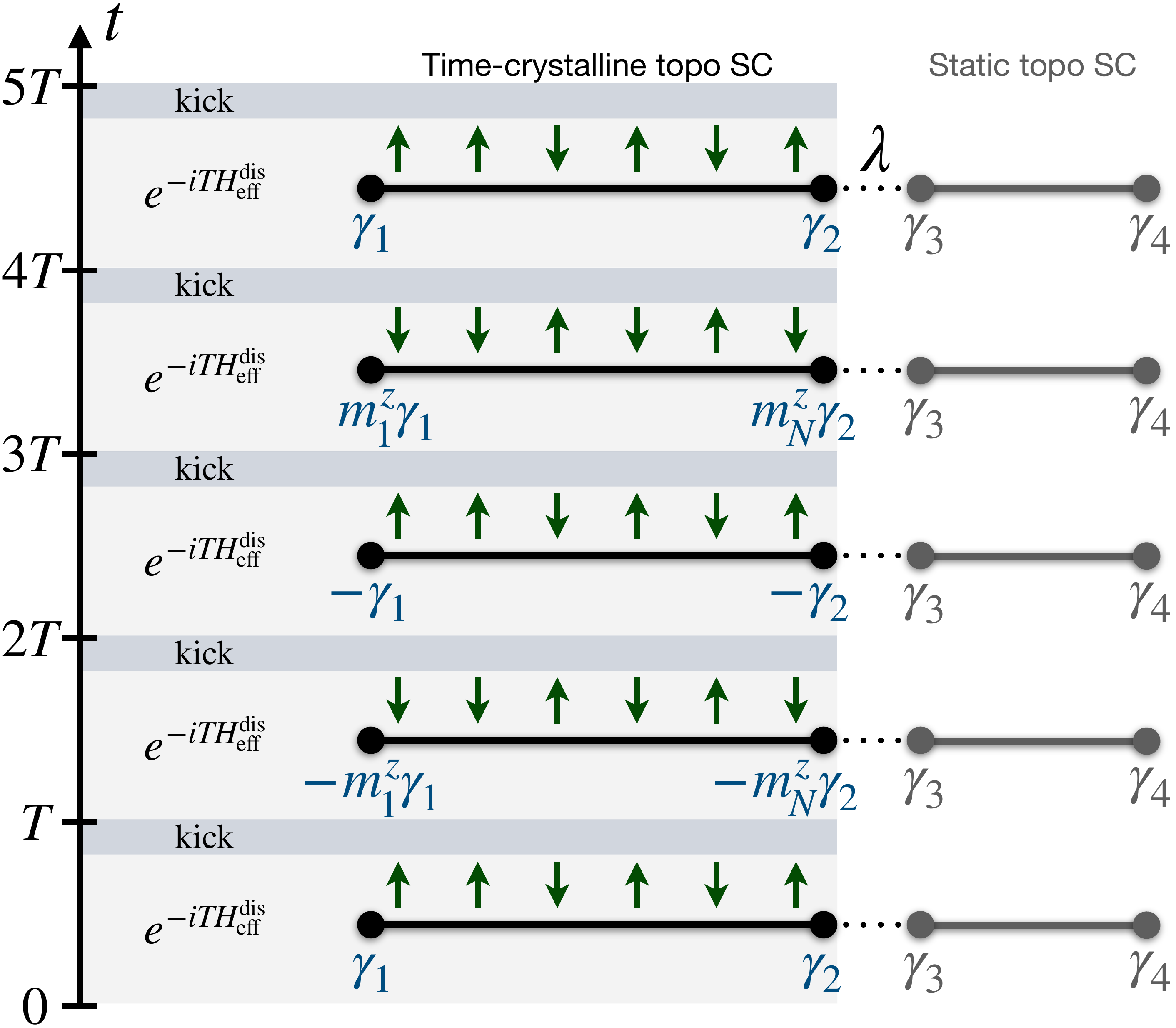}
\caption{Time evolution for the time-crystalline topological superconductor generated by Eq.~\eqref{UT} at $\epsilon = 0$.  Each period $T$ globally flips all Ising spins, yielding doubled-periodicity bulk response, whereas the Floquet Majorana modes $\gamma_{1,2}$ exhibit quadrupled-periodicity response that can be probed in the junction with the static topological superconductor on the right.  The inner Majorana modes $\gamma_{2,3}$ hybridize with coupling strength $\lambda$.  Since $\gamma_3$ is static while $\gamma_2$ evolves nontrivially after each period $T$, the junction's energy inherits the latter's quadrupled periodicity.} \label{JunctionFig}
\end{figure}

{\bf \emph{Time-crystalline topological superconductivity and detection.}}~We now promote the adiabatic ground-state phenomenon described above to a dynamic phenomenon applicable to \emph{arbitrary} physical states. To this end we apply a variation of the preceding cycle periodically with period $T$, thus generating time-crystalline topological superconductivity. 
We specifically consider a binary drive such that the Floquet operator that evolves the system over a single period reads 
\begin{align}
U_T &= e^{-i (\pi/2-\epsilon)\sum_j(m^x_j + c_j^\dagger c_j)}e^{-i T H_{\rm eff}^{\rm dis}}.
  \label{UT}
\end{align}  
The right exponential evolves the system with respect to a 
disordered, static Hamiltonian $H_{\rm eff}^{\rm dis}$ that is the same as Eq.~\eqref{StaticHamiltonian} but with $J,a,b$ replaced with random site-dependent couplings $J_j, a_j,b_j$.  
For simplicity we neglect randomness in the phases of $a_j,b_j$ and treat $J_j, a_j, b_j$ as independent random variables with magnitudes drawn from uniform distributions $[\bar J - \delta J, \bar J + \delta J],[\bar a - \delta a, \bar a + \delta a], [\bar b - \delta b, \bar b + \delta b]$.
Disorder crucially introduces many-body localization (MBL) into the dynamics and prevents heating to infinite temperatures \cite{HuseMBL,LazaridesMBL,AlessioMBL,PonteMBL,AbaninMBL}.  
The left exponential in Eq.~\eqref{UT} performs
an instantaneous `kick' that (at least approximately) flips the Ising spins via a transverse magnetic field pulse and applies a potential to the spinless fermions---thereby mimicking evolution from our adiabatic cycle without the adiabaticity requirement. 

The dynamics is analytically tractable at $\epsilon = 0$ and when $H^{\rm dis}_{\rm eff}$ reduces to Eq.~\eqref{EffHamiltonian} with random couplings $J_j, \kappa_j$. 
Starting from any Ising configuration, the `perfect' kick in $U_T$ sends $m^z_j \rightarrow -m^z_j$ and thus flips all spins, signifying period-doubling time crystallinity in the spin sector.  
In the fermionic sector, $\gamma_{1,2}$ in Eq.~\eqref{gamma_def} continue to commute with $H^{\rm dis}_{\rm eff}$ despite the randomness.  
The kick, however, nontrivially transforms the Majorana edge operators so that $U_T \gamma_1 U_T^\dagger = m^z_1 \gamma_1$ and $U_T \gamma_2 U_T^\dagger = m^z_N\gamma_2$.
Precisely as illustrated in Eq.~\eqref{FourAdiabaticCycles}, $\gamma_{1,2}$ therefore require \emph{four} drive periods to recover their initial form, i.e., they form the hallmark quadrupled-periodicity Floquet Majorana modes of the time-crystalline topological superconductor.  
Shaded regions of Fig.~\ref{JunctionFig} summarize the evolution.  

Quadrupled periodicity can be experimentally probed in junctions between time-crystalline and static topological superconductors as in the right side of Fig.~\ref{JunctionFig}, wherein $\gamma_3$ and $\gamma_4$ denote time-independent Majorana zero modes.  
Electron tunneling across the junction couples $\gamma_2$ with $\gamma_3$, producing a Hamiltonian term $H_{23} = i\lambda \gamma_2\gamma_3$ for some $\lambda$ that may depend on the adjacent Ising spins.  
Consequently, the junction's energy density (among other local properties) directly manifests the quadrupled-periodicity built into the anomalous Floquet Majorana mode $\gamma_2$.  

Rigidity against `imperfect' drives is a crucial feature of time-crystalline phases \cite{KhemaniGlass, ElseCrystal, KeyserlingkGlass, YaoCrystal}.  Here, such imperfection arises from taking $\epsilon \neq 0$ and $H^{\rm dis}_{\rm eff}$ away from the zero-correlation-length limit, which spoils exact solvability and prompts us to turn to numerics.

{\bf \emph{Numerics.}}~We employ time-evolving block decimation (TEBD), using a maximum bond dimension of $\chi = 50$, on a 20-site system with random Ising spins and parameters appropriate for our quantum-dot setup: $\phi_a = \phi_b = \pi/8$, $\bar b = \bar a/2$, $\bar J = \bar a/4$, $\mu' = 0$, $\delta a = \delta b = \delta J = \bar a/8$.  
Our simulations additionally incorporate a decoupled, static zero-energy fermion $c_0$ that functions similarly to the static topological superconductor in Fig.~\ref{JunctionFig} and allows us to numerically probe the anomalous Floquet Majorana modes with minimal computational overhead.  
For initialization, we perform a Jordan-Wigner transformation to map the fermions to bosonic spin variables and then prepare a random product state that, in fermionic language, entangles the static fermion with the rest of the system.  
We simulate the Floquet operator in Eq.~\eqref{UT} with $\bar a T = 2$ and $\bar a T = 0.2$, and with the kick shifted away from commensurability by $\epsilon = 0.2$.  
We used the ITensor Library to perform the time evolution \footnote{Calculations were performed using the ITensor Library, http://itensor.org.}.
Despite the rather small system size, in both cases the bond dimension quickly saturated, and the truncation error was relatively coarse.
To check robustness of our numerics we repeated the computations for maximum bond dimension $\chi = 25$, and the results agreed with those at $\chi = 50$.

Over a run of $60$ Floquet evolutions and $150$ disorder averages, we measure the Ising spin $\langle m^z_{j = 10}\rangle$ in the middle of the system as well as $\langle c_0^\dagger c_1\rangle$, where $c_1$ corresponds to the leftmost quantum dot.  
The former probes bulk time crystallinity while the latter probes the Floquet Majorana modes.
Figure~\ref{FreeFermionFig} plots the Fourier transform of both quantities as a function of frequency $\omega$ normalized by $\Omega = 2\pi/T$.
For $\bar a T = 2$ the data show the rigidity characteristic of a time crystal: despite the imperfect drive, the bulk magnetization and edge fermion bilinear respectively remain peaked at $\omega = \Omega/2$ and $\omega = 3\Omega/4$ (as expected for doubled-periodicity Ising spins and quadrupled periodicity Floquet Majorana modes).  
By contrast, in our $\bar a T = 0.2$ simulations both peaks clearly shift due to non-zero $\epsilon$, indicating an absence of rigid time-crystallinity for this case.  
We also ran exact numerics on a $7$-site system and measured the level-spacing statistics of the $U_T$ eigenvalues.
At $\bar a T =  2$ the mean level spacing was approximately $0.39$, close to the Poisson value $0.386$ expected for MBL \cite{PalMBL}.

\begin{figure}
\includegraphics[width = \columnwidth]{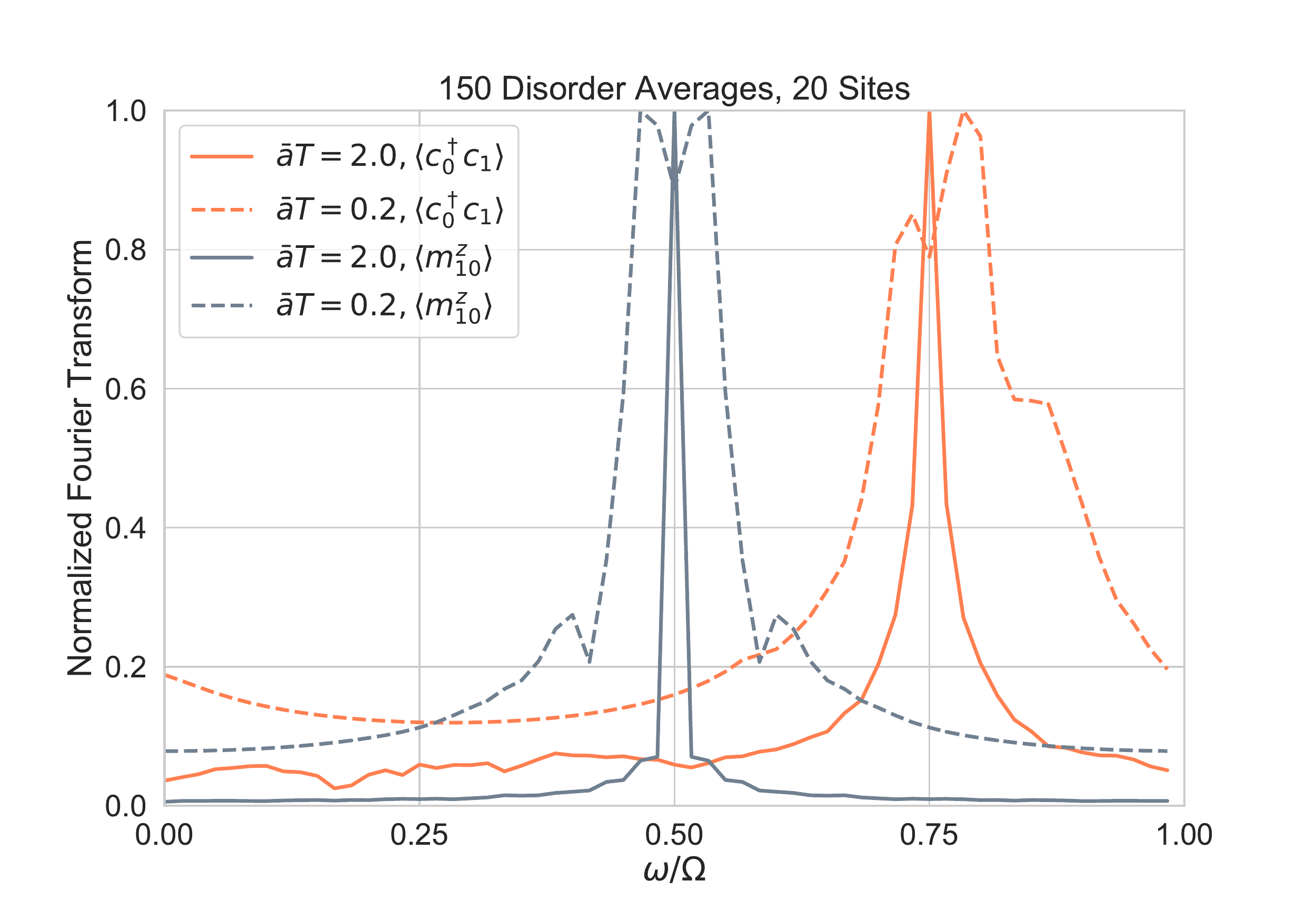}
\caption{Fourier transform of the quantities shown in the legend following time evolution via Eq.~\eqref{UT} with $\epsilon = 0.2$ and parameters specified in the main text.  Data are normalized by setting the maximum of each Fourier spectrum to 1, and frequency $\omega$ on the horizontal axis is normalized by $\Omega = 2\pi/T$, with $T$ the drive period.  
Here $m^z_{10}$ represents an Ising spin at the center of the chain, $c_0$ is an auxiliary zero-energy static fermion that enables probing the Floquet Majorana mode periodicity, and $c_1$ is the fermion at the left end of the quantum-dot chain.  
For initialization we use random Ising configurations and random fermionic states that entangle $c_0$ with the rest of the system.  
Runs were repeated $150$ times for disorder averaging with maximum bond dimension $\chi = 50$; similar results were obtained with $\chi = 25$.  
For $\bar a T = 2$ sharp peaks persist at $\Omega/2$ and $3\Omega/4$---despite `imperfect' driving generated by $\epsilon \neq 0$---indicating `rigid' doubled-periodicity Ising spins and quadrupled-periodicity Floquet Majorana modes characteristic of time-crystalline topological superconductivity.  
For $\bar a T = 0.2$, the imperfect drive pushes the peak frequencies away from these quantized values, indicating a loss of rigid time crystallinity.
}
\label{NumericsFig}
\end{figure}

{\bf \emph{Discussion.}}~The admixture of symmetry breaking and topology is known to generate new physics in static systems; examples include $8\pi$-periodic Josephson effects \cite{ZhangJosephson,OrthJosephson} and enrichment of Majorana braiding and fusion \cite{ParafermionPaper}.  
Our work establishes that driven systems can be similarly enriched by `decorating' topological phases with spontaneous time-translation symmetry breaking. 
We specifically showed that 1D time-crystalline topological superconductors engineered from quantum-dot arrays host novel Floquet Majorana modes that display anomalously long periodicity not possible with free fermions.
Exotic states of this type are not captured by the cohomology classification of interacting topological Floquet phases \cite{ElseClassification,PotterClassification,KeyserlingkClassification}.
In future work, it would be interesting to explore similarly enriched \emph{two-dimensional} (2D) phases.  
Driven spinless 2D $p+ip$ superconductors also support doubled-periodicity Floquet Majorana modes \cite{Wang2D,Yang2D,Liu2D} and thus constitute natural candidate platforms. 
One could envision promoting spinless fermions in such systems to spinful fermions coupled to magnetic degrees of freedom as done here, possibly leading to new higher-dimensional adiabatic cycles, time-crystalline topological phases, and nontrivial pre-thermal regimes. 

{\bf \emph{Acknowledgments.}}~It is a pleasure to thank David Weld and Norm Yao for illuminating discussions.  
This work was supported by the Army Research Office under Grant Award W911NF-17-1-0323; 
the NSF through grant DMR-1723367; 
grant No. 2016258 from the United States-Israel Binational
Science Foundation (BSF);  the Israel Science Foundation (ISF);
the Caltech Institute for Quantum Information and Matter, an NSF Physics Frontiers Center with support of the Gordon and Betty Moore Foundation through Grant GBMF1250; 
the Walter Burke Institute for Theoretical Physics at Caltech; and the Gordon and Betty Moore Foundation's EPiQS Initiative, Grant GBMF8682 to JA.

\bibliography{TC_references}
\setcounter{secnumdepth}{2}

\appendix

\section{Derivation of effective spinless-fermion Hamiltonian}
\label{ProjectionAppendix}

Here we derive the effective Hamiltonian given in Eq.~\eqref{StaticHamiltonian} that describes the quantum dots and Ising spins in the limit of large $K$.  
We start from the original microscopic model $H = H_0 + H_f$ [recall Eq.~\eqref{H}] and decompose the spinful fermions via
\begin{align}
  f_{j\uparrow} &= \frac{1}{2}[(1+m^z_j) c_j + (1-m^z_j) d_j] \\
  f_{j\downarrow} &= \frac{1}{2}[(1-m^z_j) c_j + (1+m^z_j)d_j]. 
\end{align}
Here $c_j$ are precisely the low-energy fermionic degrees of freedom from Eq.~\eqref{cj} that minimize the energy of the $K$ term, while $d_j$ represent high-energy fermions that we wish to formally integrate out.  
In terms of $c_j$ and $d_j$, we have
\begin{align} 
  H_0 &= \sum_j[-J m^z_j m^z_{j+1} - K (c_j^\dagger c_j-d_j^\dagger d_j)]
\end{align}
and
\begin{align}
  H_f &= \sum_j \big{\{}-\mu (c_j^\dagger c_j + d_j^\dagger d_j) 
  \nonumber \\
  &+  [(-t P_{j1} + i\alpha P_{j2}) (c_j^\dagger c_{j+1} + d_j^\dagger d_{j+1})  + H.c.]
  \nonumber \\
&+ [(-t P_{j2} + i\alpha P_{j1}) (c_j^\dagger d_{j+1} + d_j^\dagger c_{j+1}) + H.c.]
 \nonumber \\
&+ \Delta m^z_j (c_j d_j + H.c.) \big{\}}
\label{EffHamiltonian2}.
\end{align}
In Eq.~\eqref{EffHamiltonian2} we introduced projectors
\begin{align}
P_{j1} = \frac{1}{2}(1 + m^z_j m^z_{j+1}),~~~
P_{j2} = \frac{1}{2}(1 - m^z_j m^z_{j+1})
\end{align} 
that project onto states where nearest-neighbor Ising spins are aligned and anti-aligned, respectively.

The formal elimination of $d_j$'s is conveniently carried out within a (Euclidean) path-integral formalism, with the zero-temperature partition function given by
\begin{equation}
  Z = \int \mathcal{D}d^\dagger \mathcal{D}d\mathcal{D}c^\dagger \mathcal{D}c e^{-S},
\end{equation}
where 
\begin{equation}
  S = \int_{-\infty}^\infty d\tau\left[\sum_j(c_j^\dagger \partial_\tau c_j + d_j^\dagger \partial_\tau d_j) + H\right]
\end{equation}
is the imaginary-time action.
Upon integrating over $d_j,d_j^\dagger$ (which can be done exactly since $H$ is quadratic in fermions), the partition function can be written as
\begin{align}
  Z &\propto \int \mathcal{D}c^\dagger \mathcal{D}c e^{-S_{\rm eff}}
  \nonumber \\
  S_{\rm eff} &= \int_{-\infty}^\infty \frac{d\omega}{2\pi}\left[\sum_j (-i \omega c_j^\dagger c_j) + \mathcal{H}_{\rm eff}(\omega) \right].
\end{align}
In the low-frequency limit, i.e., $|\omega| \ll (K-\mu)$, we can neglect frequency dependence in $\mathcal{H}_{\rm eff}$ to obtain an effective spinless-fermion Hamiltonian that takes the form of Eq.~\eqref{StaticHamiltonian}.   
Finally, upon truncating the chemical potential, hopping, and pairing matrix elements to leading nontrivial order in $1/(K-\mu)$, we obtain precisely the $\mu', t'_{m^z_j,m^z_{j+1}}, \Delta'_{m^z_j,m^z_{j+1}}$ couplings quoted in the main text.  

\section{Transfer-matrix details}
\label{TransferMatrixApp}

To examine the fermionic ground state for random Ising spins we express the model of Eq.~\eqref{StaticHamiltonian} in terms of transfer matrices. 
The equation of motion for $\psi_j =(c_j,c_j^\dagger)$ can be brought to the form
\begin{align}
 \begin{pmatrix} \psi_{j+1} \\ F_j^\dagger \psi_j\end{pmatrix} &=T_j
 \begin{pmatrix} \psi_{j} \\ F_{j-1}^\dagger \psi_{j-1}\end{pmatrix}~,
\end{align}
with
\begin{align}
T_j &= \begin{pmatrix}F_{j}^{-1}[E - \mu\sigma^z] &-F_{j}^{-1}\\F_j^\dagger &0\end{pmatrix}~,\\
F_{j} &=
 \begin{pmatrix}t'_{m_j^z,m_{j+1}^z} &-\Delta^{\prime *}_{m_j^z,m_{j+1}^z}\\\Delta^{\prime }_{m_j^z,m_{j+1}^z} &-t^{\prime *}_{m_j^z,m_{j+1}^z}~\end{pmatrix}.
\end{align}

\begin{figure}
\includegraphics[width =.49\columnwidth]{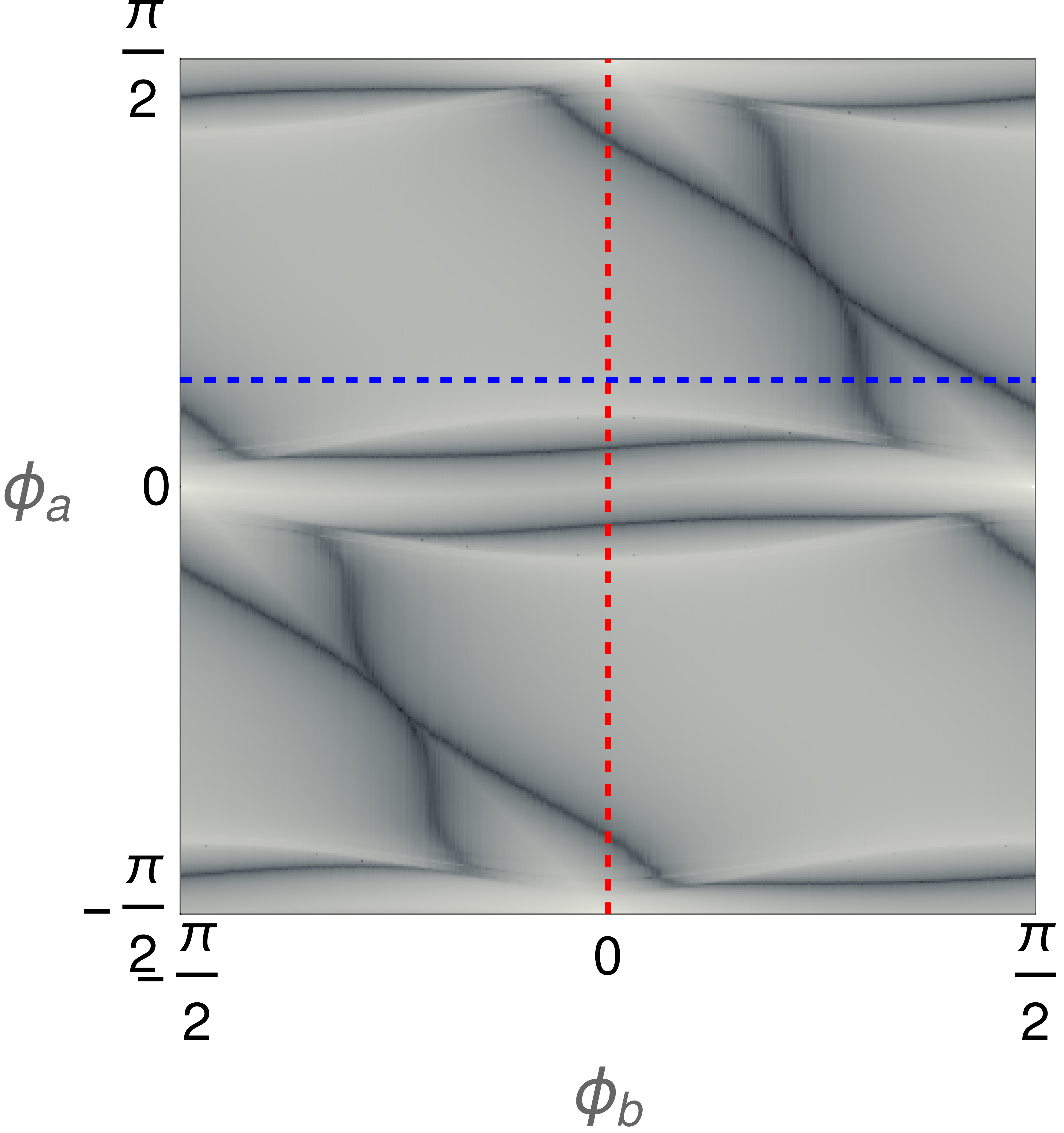}
\includegraphics[width =.49\columnwidth]{cuts.png}
\caption{Transfer-matrix data for $\mu' =| b| = |a|/4$ and $10^6$ sites. On the left we show a density map of $\log(\xi)$, with darker shades denoting larger $\xi$. The phase boundaries are readily apparent as narrow dark lines. The dashed lines denote two specific cuts for which we show $\xi$ on a linear scale on the right. The very rapid divergence of $\xi$ near specific points supports our identification of the phase boundaries.}
\label{fig-2525}
\end{figure}

The transfer matrix for an $N$-site chain is $Q = \prod_{j=1}^N T_j$, and the smallest positive eigenvalue of $\frac{1}{N}\log[Q Q^\dagger]$ is the inverse localization length $\xi^{-1}$ (see, e.g., \cite{MacKinnon2003}). In Fig.~\ref{fig-2525} we present the data from which the phase diagram in Fig.~\ref{FreeFermionFig} of the main text is obtained. First, we show a two-dimensional density map of $\xi$ on a logarithmic scale, which reveals the phase boundaries without any need for processing the data. Second, we show $\xi$ on a linear scale for two representative cuts to illustrate the rapid growth of $\xi$ near phase boundaries. 

\begin{figure}
\includegraphics[width =.32\columnwidth]{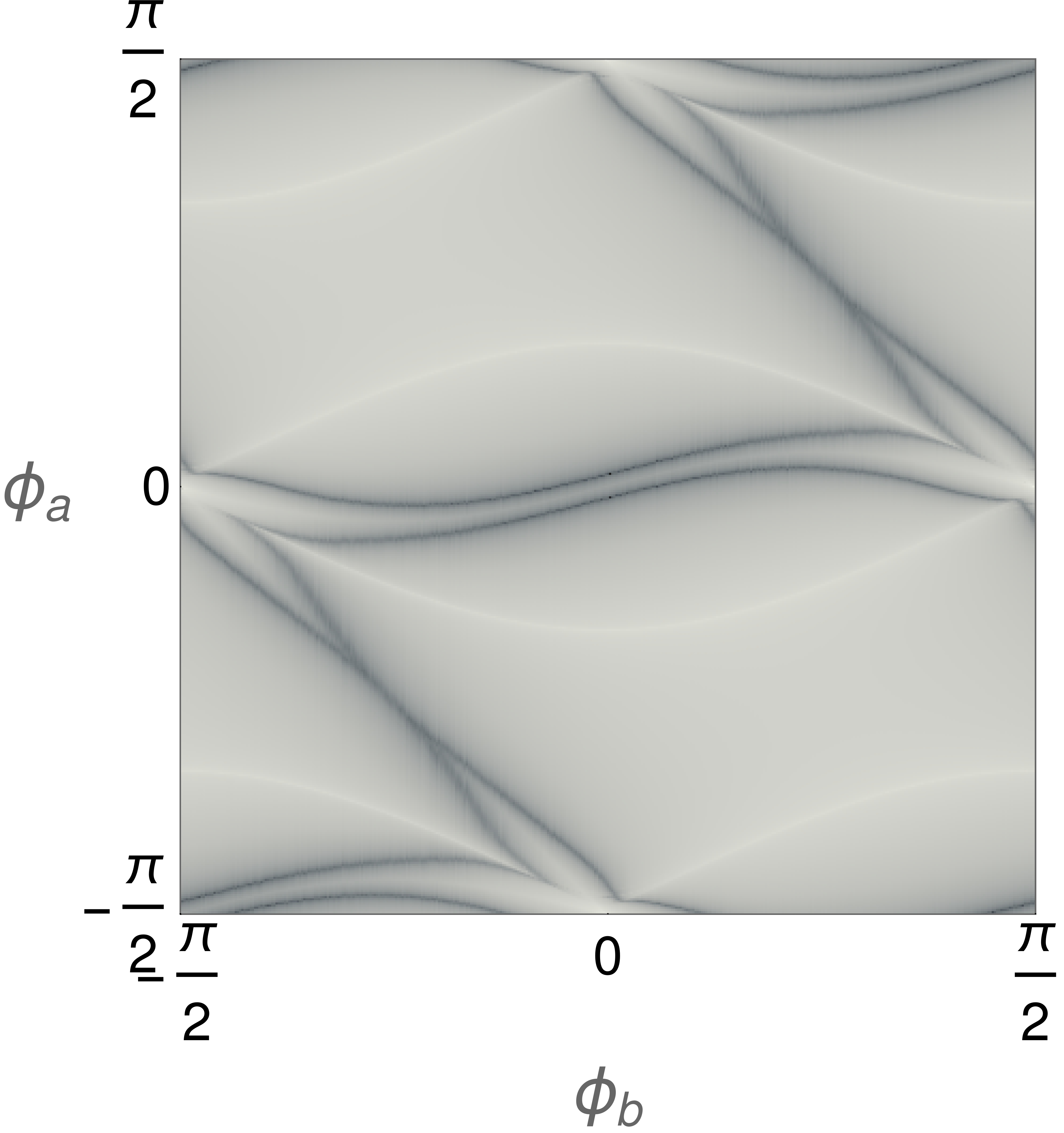}
\includegraphics[width =.32\columnwidth]{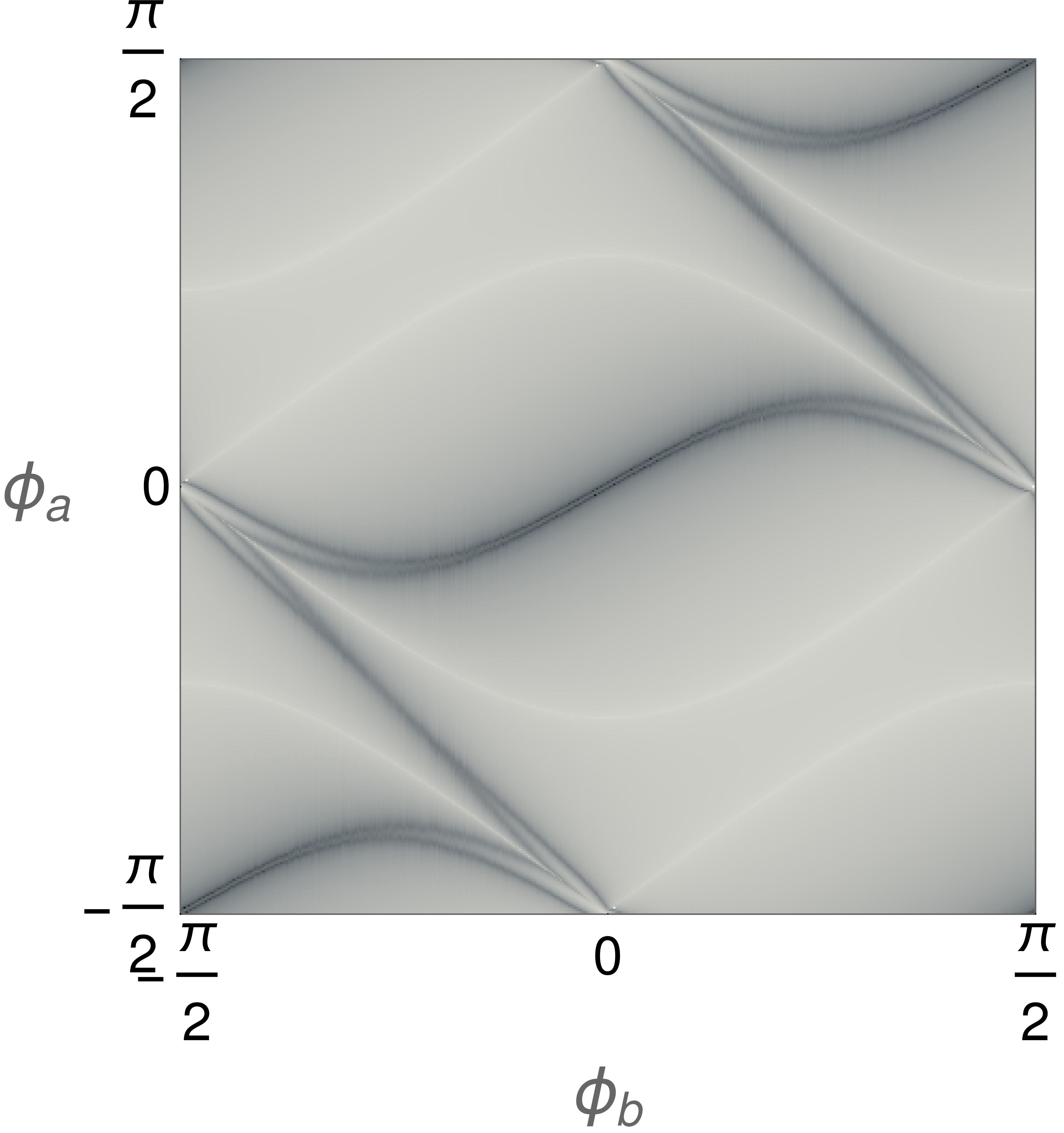}
\includegraphics[width =.32\columnwidth]{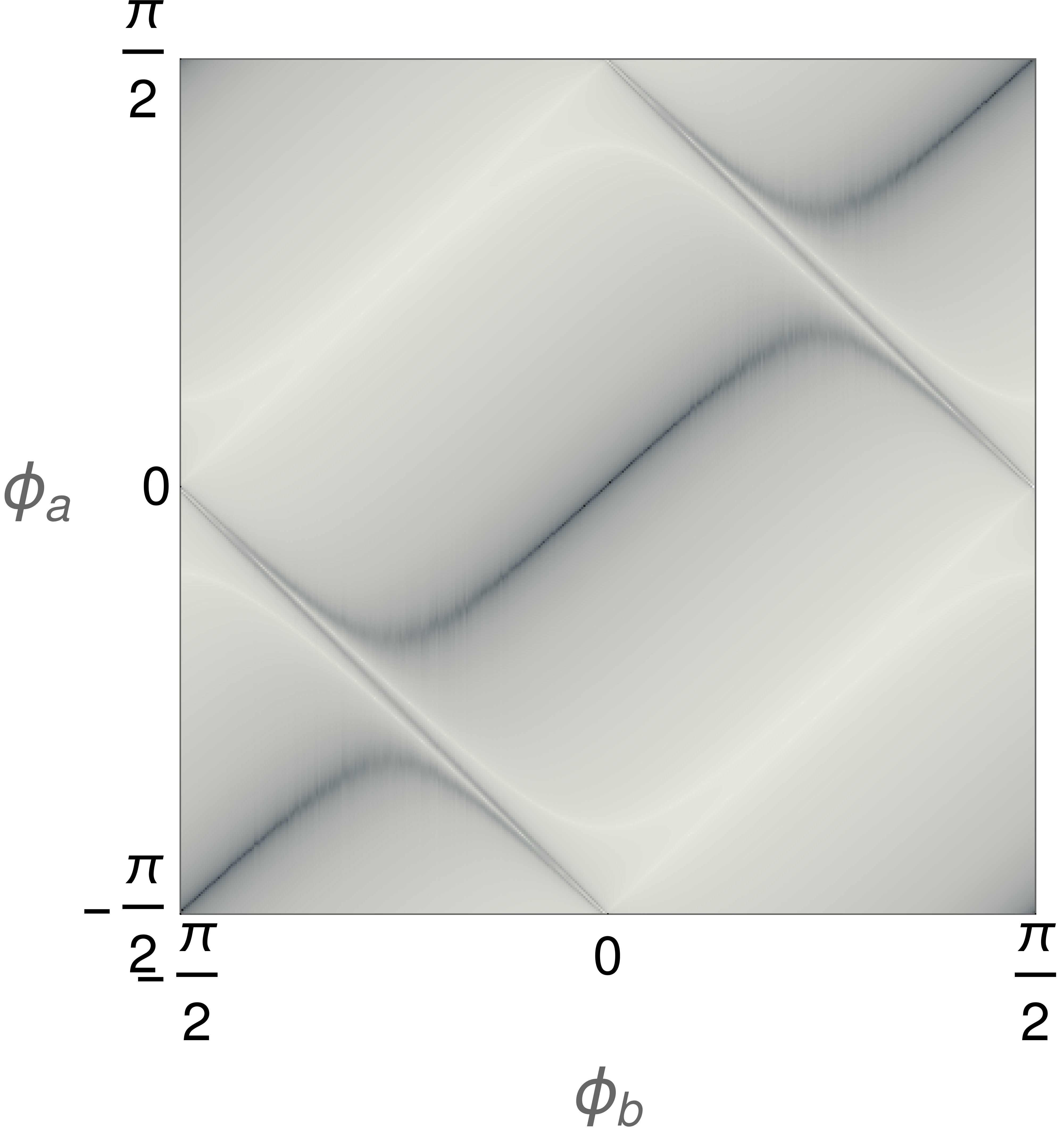}
\caption{Density maps of $\log(\xi)$ for (left) $\mu' = |a|/8$ and $|b|=|a|/2$, (middle) $\mu' = |a|/20$ and $|b|=3|a|/4$, and (right) $\mu' = |a|/50$ and $|b|=0.95|a|$. The phase boundaries change significantly between these parameter values, but the special point $\phi_a = -\phi_b = \pi/4$ always remains deeply in the localized topological phase.}
\label{fig-evol}
\end{figure}

Finally, we detune the parameters of the models from the ones of Fig.~\ref{FreeFermionFig}---which relate to the microscopic model---towards the exactly solvable point $|a| = |b|$ and $\mu'=0$; see Fig.~\ref{fig-evol}. [In Figs.~\ref{fig-2525} and \ref{fig-evol} we do not use the relation $\xi(\phi_a,\phi_b) = \xi(-\phi_a,-\phi_b)$ to halve the data points, contrary to Fig.~\ref{FreeFermionFig}(b) from the main text.] During this deformation the phase boundaries move substantially, but at the specific value  $\phi_a = -\phi_b = \pi/4$ the system always remains in the same strongly localized topological phase. Consequently, the topological phase obtained with microscopically derived parameters indeed smoothly connects to the zero-correlation length limit $\phi_a = -\phi_b = \pi/4$, $|a| = |b|$, and $\mu' = 0$ as suggested by Fig.~\ref{FreeFermionFig}(b) from the main text.  

\section{Majorana Zero Modes via the Born Approximation}
\label{BornAppendix}

For certain values of $a,b$, we can use the Born approximation to capture Majorana zero modes in the Hamiltonian of Eq.~\eqref{StaticHamiltonian} with random $m^z_j$ Ising configurations.
In what follows we ignore the $J$ term for simplicity.  
Suppose that we perform the gauge transformation
\begin{align}
c_j &\rightarrow e^{i \frac{\pi}{4}(1-m^z_1)}e^{-i\frac{\pi}{4} [1 + \sum_{k < j} (1 - m^z_{k} m^z_{k+1}) ]} c_j,
\label{GaugeTransform}
\end{align}
so that Eq.~\eqref{StaticHamiltonian} becomes
\begin{align}
H_{\rm eff} &= \sum_j [-\mu' c_j^\dagger c_j 
\nonumber \\
 &+ (t''_{m_j^z,m_{j+1}^z} c_j^\dagger c_{j+1}  + \Delta''_{m_j^z,m_{j+1}^z} c_j c_{j+1} +H.c.)].
\label{StaticHamiltonian2}
\end{align} 
The new hopping and pairing coefficients are given by
\begin{align}
  t''_{m_j^z,m_{j+1}^z} &= \frac{(a e^{-i\frac{\pi}{4}} + c.c.)}{\sqrt{2}} + \frac{m^z_j m^z_{j+1}(a e^{i \frac{\pi}{4}} + c.c.)}{\sqrt{2}}
  \\
  \Delta''_{m_j^z,m_{j+1}^z} &= \frac{(-b e^{i\frac{\pi}{4}} + c.c.)}{\sqrt{2}} + \frac{m^z_j m^z_{j+1}(b e^{-i \frac{\pi}{4}} + c.c.)}{\sqrt{2}}.
\end{align}
As before we write $a = |a| e^{i \phi_a}$ and $b = |b| e^{i \phi_b}$.  
Notice that at $\phi_a = \pi/4$ and $\phi_b = -\pi/4$, which are the same phases used to access the zero-correlation limit, \emph{the $m^z_j$ dependence has been completely gauged out of the Hamiltonian for any $|a|,|b|$}.
We immediately conclude that at these phases the system harbors edge Majorana zero modes regardless of the Ising configuration provided $|\mu'| < 2 \sqrt{2}|a|$.  

Suppose next that we deform away from this limit by writing $\phi_a = \pi/4 + \epsilon_a$ and $\phi_b = -\pi/4 + \epsilon_b$, where $|\epsilon_{a,b}| \ll 1$.  
The $m^z_j$ dependence no longer drops out, and for random Ising configurations can be viewed as generating weak disorder in the fermion hoppings and pairings.  
To lowest order in the Born approximation this disorder is treated by simply replacing $H_{\rm eff} \rightarrow \overline{H_{\rm eff}}$ with the overline indicating a disorder average over $m^z_j$ configurations.  
Here and below we will assume that the $m^z_j$'s are uncorrelated from site to site and have zero mean (as appropriate for the random Ising configurations that are our primary interest).  
The hopping and pairing strengths accordingly become 
\begin{align}
  \overline{t''_{m^z_j,m^z_{j+1}}} = \sqrt{2} |a|\cos \epsilon_a, ~~\overline{\Delta''_{m^z_j,m^z_{j+1}}} = - \sqrt{2} |b|\cos \epsilon_b.
\end{align}
Within this approximation edge Majorana zero modes persist so long as $|\mu'| < 2 \sqrt{2}|a|\cos \epsilon_a$.  

Thus far we have made no assumptions about the relative strength of $|a|$ and $|b|$.  
Additional progress is possible if we specialize to the (most physically relevant) regime $|a| \gg |b|$, which we now assume.  
We continue to take $\phi_a = \pi/4 + \epsilon_a$ but now allow for general $\phi_b$, and treat $\epsilon_a$ as well as the entire pairing term as perturbations.    
Within the lowest-order Born approximation the hopping and pairing strengths are now modified to
\begin{align}
  \overline{t''_{m^z_j,m^z_{j+1}}} &= \sqrt{2} |a|\cos \epsilon_a, 
  \label{tpp} \\
  \overline{\Delta''_{m^z_j,m^z_{j+1}}} &= -\sqrt{2} |b|\cos(\phi_b+\pi/4).
  \label{Deltapp}
\end{align}
At this order, edge Majorana zero modes appear when $|\mu'| < 2 \sqrt{2}|a|\cos \epsilon_a$ \emph{and} $\cos(\phi_b + \pi/4) \neq 0$.  
These criteria naively rule out Majorana zero modes when $\phi_b = \pi/4$.  
Nonzero pairing after disorder averaging is, however, generated at \emph{second order} in the Born approximation (at least when $\epsilon_a \neq 0$), so that Majorana zero modes can still emerge as we show next.

Let $\phi_b = \pi/4$ and write the Hamiltonian as $H_{\rm eff} = H_0 + H_1$, where all $m^z_j$-dependent terms are lumped into $H_1$.  
Explicitly, we have
\begin{align}
  H_0 &=\sum_j [-\mu' c_j^\dagger c_j + (\bar t c_j^\dagger c_{j+1}+H.c.)]
  \\
  H_1 &= \sum_j m^z_j m^z_{j+1}(t_1 c_j^\dagger c_{j+1} + \Delta_1 c_j c_{j+1} + H.c.)
\end{align}
with $\bar t = \sqrt{2}|a| \cos\epsilon_a$, $t_1 = -\sqrt{2}|a|\sin\epsilon_a$, and $\Delta_1 = \sqrt{2}|b|$.
To proceed we switch to first-quantized language, defining position-space Hamiltonian matrix elements ${\cal H}_{0,1;jk}$ through
\begin{align}
H_{0,1} = \sum_{j,k} \Psi^\dagger_j  {\cal H}_{0,1;jk} \Psi_k,
\end{align}
where
\begin{align}
\Psi^\dagger_j = \begin{bmatrix}
c^\dagger_j & c_j
\end{bmatrix}
\end{align}
is the Nambu spinor.
In terms of the bare Green's function
\begin{equation}
  G_{0;jk}(i\omega) = (i\omega - {\cal H}_0)_{jk}^{-1},
\end{equation}  
the fermion self-energy at second order in the Born approximation reads
\begin{equation}
  \Sigma_{jk} = \overline{{\cal H}_{1;jl} G_{0;lm}(i\omega = 0) {\cal H}_{1;mk}}.
\end{equation} 
Repeated indices are implicitly summed above.  
The prefactor $m^z_j m^z_{j+1}$ in $\mathcal{H}_1$ implies that the disorder average is nonzero only when we contract matrix elements corresponding to the same sites, i.e., when $jl = mk$ or $jl  = km$.

Disorder averaging effectively restores translation invariance, so it is useful to pass to momentum space.
For $\mathcal{H}_0$ we simply write
\begin{align}
  \mathcal{H}_{0;jk} &= \int_p e^{i p(j-k)} {\mathcal{H}}_{0}(p).
\end{align}
The Fourier transform is 
\begin{equation}
    {\cal H}_0(p) = \frac{1}{2}(2 \bar t \cos{p} - \mu') \tau^z,
\end{equation} 
where Pauli matrices $\tau^{x,y,z}$ act in Nambu space.
For $\mathcal{H}_1$ we isolate the position-dependent magnetization by instead writing
\begin{align}
  {\cal H}_{1;jk} = m^z_j m^z_{k} \int_{p} e^{ip(j-k)} \tilde {\cal H}_1 (p),
\end{align}
which yields
\begin{align}
\tilde {\cal H}_1(p) &=  t_1 \cos{p} ~\tau^z + \Delta_1 \sin{p} ~\tau^y.
\end{align} 
We can now express the self-energy as
\begin{align}
&\Sigma_{jk} = \overline{m^z_j m^z_l m^z_m m^z_k} 
\nonumber \\
&\times \int_{p_1,p_2} e^{ip_1(j-l)}e^{ip_2(m-k)} \tilde {\cal H}_1 (p_1) G_{0;lm}(i\omega = 0) \tilde {\cal H}_1 (p_2).
\end{align}
The disorder average on the first line evaluates to
\begin{align}
 \overline{m^z_j m^z_l m^z_m m^z_k}  = \delta_{jk} \delta_{lm} + \delta_{jm} \delta_{kl}.
\end{align}
The first pair of Kronecker deltas involve $\delta_{jk}$ and thus merely generate an on-site correction. 
We neglect this term and instead focus on the second pair of Kronecker deltas:
\begin{equation}
  \Sigma_{jk} \rightarrow \int_{p_1,p_2} e^{i(p_1+p_2)(j-k)} \tilde {\cal H}_1 (p_1) G_{0;kj}(i\omega = 0) \tilde {\cal H}_1 (p_2).
\end{equation}
Upon further Fourier transforming the Green's function we obtain
\begin{align}
  \Sigma_{jk} &= \int_q e^{i q(j-k)} \Sigma(q)
  \\
  \Sigma(q) &= \int_{p_1,p_2} \tilde {\cal H}_1 (p_1) G_0(i\omega = 0,p_1+p_2-q) \tilde {\cal H}_1 (p_2).
\end{align}
It is useful to now decompose the self energy as
\begin{equation}
  \Sigma(q) = \Sigma^z(q) \tau^z + \Sigma^y(q) \tau^y.
\end{equation}
The $\Sigma^z(q)$ part encodes renormalization of the kinetic energy, while $\Sigma^y(q)$ encodes $p$-wave pairing.  
The latter is given by
\begin{align}
  \Sigma^y(q) &= -2t_1 \Delta_1 \int_{p_1,p_2} \frac{\sin(p_1+p_2)}{2 \bar t\cos(p_1+p_2-q)-\mu'}
  \\
  &= -\frac{t_1 \Delta_1}{\bar t} f\left(\frac{\mu'}{2\bar t}\right) \sin q
  \label{Sigmay}
\end{align}
for some nontrivial function $f(x)$ that satisfies $f(x \ll 1) \approx 1$.  
Provided $t_1, \Delta_1$ are nonzero---which in turn requires nonzero $\epsilon_a$ and $|b|$---the pairing amplitude is finite, yielding unpaired Majorana modes if $|\mu'| < 2 \sqrt{2} |a| \cos \epsilon_a$ as claimed.
We note that the correlated nature of disorder in the tunneling and pairing terms in $H_1$ is essential to this outcome.  

In our second-order Born analysis we set $\phi_b = \pi/4$ exactly.  
If we now take $\phi_b = \pi/4 + \epsilon_b$ (again with $\epsilon_b \ll 1$) then we can estimate the effective $p$-wave pairing amplitude $\Delta_{\rm eff}$ by simply summing the contributions from Eqs.~\eqref{Deltapp} and \eqref{Sigmay}. 
[Technically, taking $\epsilon_b \neq 0$ also modifies Eq.~\eqref{Sigmay}, though this correction will be small compared to the contribution from Eq.~\eqref{Deltapp}.]
We thereby obtain 
\begin{align}
  \Delta_{\rm eff} &\approx \sqrt{2} |b| \epsilon_b - \frac{t_1 \Delta_1}{\bar t} f\left(\frac{\mu'}{2\bar t}\right)
  \nonumber \\
  &\approx \sqrt{2} |b| \left[\epsilon_b + \epsilon_a f\left(\frac{\mu'}{2\sqrt{2} |a|}\right)\right]
  \label{Deltaeff}
\end{align}
where on the second line we used $t_1/\bar t \approx -\epsilon_a$, $\bar t \approx \sqrt{2} |a|$, and $\Delta_1 = \sqrt{2}|b|$.  
In the limit $\mu' \ll |a|$ we can further replace $f\rightarrow 1$; the pairing then vanishes when $\epsilon_b = -\epsilon_a$, which defines a gapless line along which Majorana modes are absent.  

\begin{figure}
\includegraphics[width = \columnwidth]{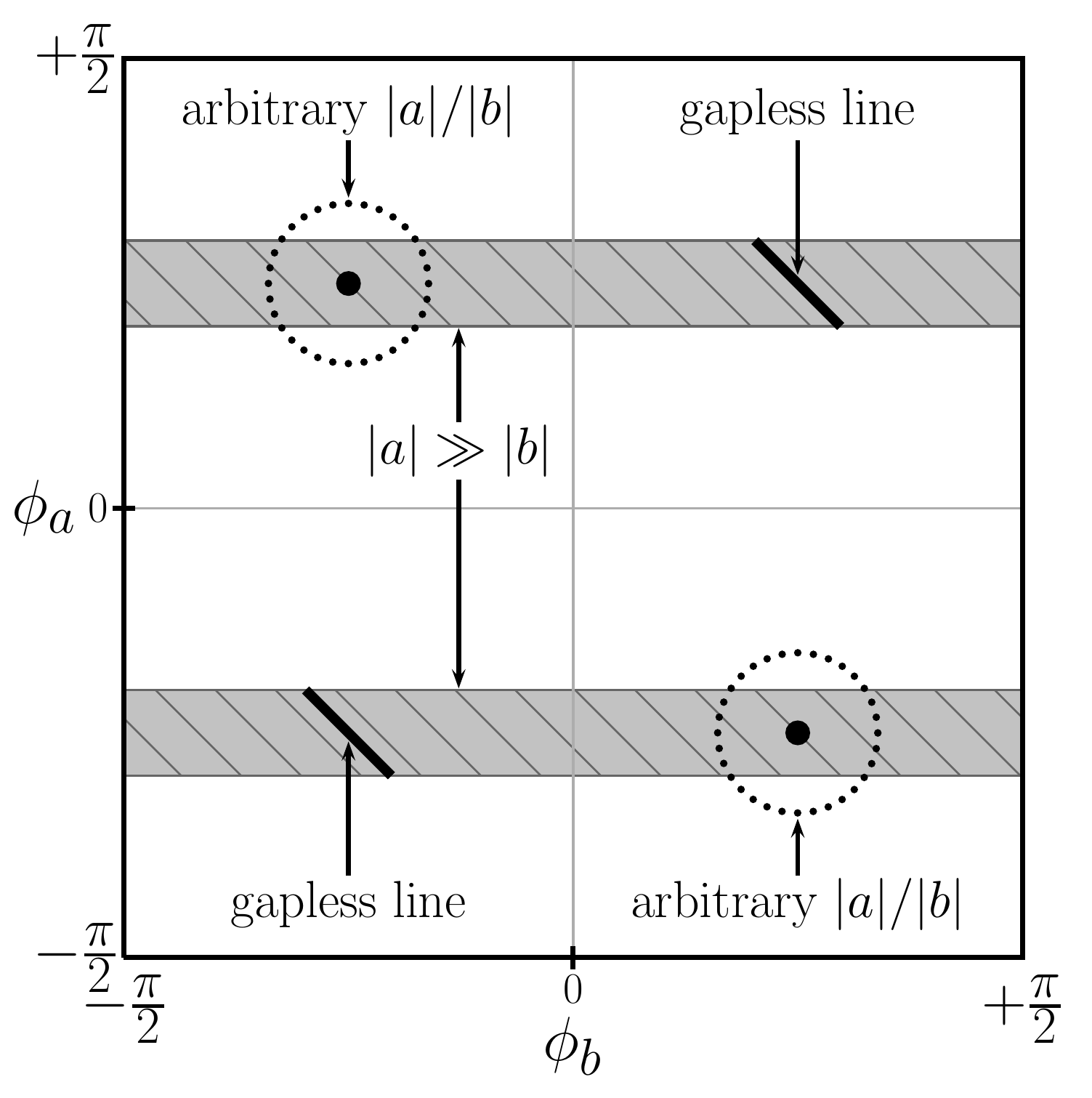}
\caption{Summary of Born-approximation results.  Shaded and circled regions denote $\phi_{a,b}$ values amenable to the Born approximation (assuming the regime of $|a|/|b|$ values indicated).   Except for the gapless lines in the upper-right and lower-left quadrants, Majorana zero modes are predicted over a finite window of chemical potential throughout these regions, in agreement with transfer-matrix simulations.   }
\label{BornFig}
\end{figure}

Figure~\ref{BornFig} summarizes our Born-approximation results, which are fully consistent with our transfer-matrix simulations.

The Born approximation further elucidates the structure of the phase diagram.
After applying the gauge transformation in Eq.~\eqref{GaugeTransform}, the Hamiltonian in Eq.~\eqref{StaticHamiltonian2} exhibits purely real couplings.
Hence an `accidental' antiunitary $\mathcal{T}'$ symmetry that obeys $\mathcal{T}'^2 = +1$ becomes manifest.  
Majorana modes can therefore be classified as `real' or `imaginary' depending on whether they exhibit eigenvalues $+1$ or $-1$ under $\mathcal{T}'$.
In the standard, uniform Kitaev chain Hamiltonian, the topological phase can be characterized by the relative sign of the hopping and pairing, $\text{sgn}(t \Delta)$.
Should this quantity be positive, the left Majorana zero mode is imaginary while its partner on the right end is real.
If the sign is negative, the opposite is true.

Our system is more complex, in that the hopping and pairings depend nontrivially on the Ising configuration in a site-dependent fashion.  
However, the Born approximation smears out this nontrivial dependence, thereby generating uniform effective hopping and pairing.  
With $|a| \gg |b|$, $\phi_a = \pi/4+\epsilon_a$, and $\phi_b = \pi/4 + \epsilon_b$, these quantities are given approximately by Eqs.~\eqref{tpp} and \eqref{Deltaeff}.   
In particular, the effective pairing in Eq.~\eqref{Deltaeff} changes sign along the gapless lines sketched in Fig.~\ref{BornFig}---implying that the two topological phases meeting at that line exhibit Majorana zero modes with opposite $\mathcal{T}'$ eigenvalues.  
More generally, a first-order or continuous phase transition, or an intermediate state, necessarily separates these phases so long as $\mathcal{T}'$ persists.  

\section{Transformation of Majorana Zero Modes}
\label{SolutionAppendix}

In the main text we deformed our effective spinless-fermion Hamiltonian to the zero-correlation-length limit, yielding Eq.~\eqref{EffHamiltonian}.  
Each Majorana zero mode in this limit localizes to a single site as shown in Fig.~\ref{SetupFig} and Eq.~\eqref{gamma_def}.
Moreover, according to Eq.~\eqref{TransformationRules} each Majorana zero mode acquires a factor of the adjacent Ising spin, i.e., $m^z_1$ or $m^z_N$, under time-reversal symmetry $\mathcal{T}$.
This transformation rule raises a conundrum: away from the zero-correlation-length limit, the zero-mode wavefunctions extend into the bulk over a distance set by the correlation length, and thus `sample' not just $m^z_1$ or $m^z_N$, but many Ising spins.  
How does $\mathcal{T}$ transform the Majorana zero modes in this more generic situation? 
The normalization $\gamma^2=1$ together with Hermiticity implies that the zero-mode operators can only be multiplied by an operator with eigenvalues $\pm 1$. 
This discreteness prohibits any perturbative corrections and the transformation in Eq.~\eqref{TransformationRules} in fact continues to hold more generally. 
It is instructive to see explicitly how this comes about by perturbing the Hamiltonian Eq.~\eqref{EffHamiltonian} away from the perfectly dimerized limit. 

Consider the $\mathcal{T}$-invariant Hamiltonian
\begin{align}
H_\text{eff}'' = \sum_j(&-J m^z_j m^z_{j+1} 
-i\kappa s_{m^z_j,m^z_{j+1}} \eta_{Aj} \eta_{Bj+1} \nonumber \\
&-i\kappa' \eta_{Aj} \eta_{Bj})
\end{align}  
corresponding to Eq.~\eqref{EffHamiltonian} modified by the $\kappa'$ term---which spoils the perfect dimerization and yields a finite correlation length. 
We assume $|\kappa'|<|\kappa|$ so that the fermions remain in the topological phase, and also take $\kappa'$ to be independent of $m^z_j$'s since such a choice is compatible with $\mathcal{T}$.  
In contrast, $\mathcal{T}$ necessitates the nontrivial $m^z$ dependence in the signs $s_{m^z_j,m^z_{j+1}}$.
This $m^z$ dependence can nevertheless be absorbed into the Majorana fermions by defining
\begin{align}
\eta_{A,B j} \equiv \left(\prod_{k<j} s_{m^z_k,m^z_{k+1}}\right) \tilde \eta_{A,B j},
\end{align} 
where $\tilde\eta_{A,B j}$ are a new set of Majorana operators.
The Hamiltonian becomes
\begin{align}
H_\text{eff}'' = \sum_j(&-J m^z_j m^z_{j+1} 
-i\kappa \tilde{\eta}_{Aj} \tilde{\eta}_{Bj+1} -i\kappa' \tilde{\eta}_{Aj} \tilde{\eta}_{Bj}).
\end{align}
Couplings between Majorana fermions in this representation are manifestly independent of the Ising spins.  

Because $H_{\rm eff}''$ only couples $\tilde \eta_{Ai}$ Majorana fermions to $\tilde \eta_{Bj}$ Majorana fermions, the Hamiltonian preserves an `accidental' antiunitary symmetry $\mathcal{T}'$ (see also Appendix~\ref{BornAppendix}) that obeys $(\mathcal{T}')^2 = +1$ 
and sends
\begin{align}
  m^z_j \rightarrow m^z_j, ~~~\tilde\eta_{Aj} \rightarrow -\tilde \eta_{Aj}, ~~~\tilde\eta_{Bj} \rightarrow \tilde\eta_{Bj}.
\end{align}  
The zero modes $\gamma_{1,2}$ can be defined such that they acquire either $+1$ or $-1$ eigenvalue under $\mathcal{T}'$, which sharply constrains their allowed form.  
Additionally incorporating Hermiticity and invoking continuity with the $\kappa' = 0$ limit allows us to write
\begin{align}
\gamma_1 &= \sum_{j} \phi_{Bj} \tilde{\eta}_{Bj} =  \sum_{j} \phi_{Bj} \left(\prod_{k<j} s_{m^z_k,m^z_{k+1}}\right) \eta_{Bj}\\
\gamma_2 &= S\sum_{j} \phi_{Aj} \tilde{\eta}_{Aj} = \sum_{j} \phi_{Aj} \left(\prod_{k\geq j} s_{m^z_k,m^z_{k+1}}\right) \eta_{Aj}
\end{align} 
for real $\phi_{A,Bj}$ that localize exponentially to the ends of the chain and, importantly, do not depend on $m^z_j$. 
On the right sides we reverted back to $\eta_{A,Bj}$ operators to explicitly display the \emph{non-local} $m^z_j$ dependence in the zero-mode wavefunctions. 
In the second line we introduced a factor $S = \prod_{{\rm all~sites~}j}  s_{m^z_j,m^z_{j+1}}$, which causes the string of $s_{m^z_k,m^z_{k+1}}$ signs to emanate from the right in the expression for $\gamma_2$.  
This convention is very natural since $\gamma_2$ localizes to the right end of the chain, and moreover correctly recovers the $\kappa' = 0$ limit of $\gamma_2$ from Eq.~\eqref{gamma_def}.

Physical time reversal $\mathcal{T}$ sends
\begin{align}
&\eta_{Aj} \rightarrow m_j^z \eta_{Aj}, ~\eta_{Bj} \rightarrow -m_j^z \eta_{Bj} ,\\
&s_{m^z_j, m^z_{j+1}} \rightarrow m^z_j m^z_{j+1} s_{m^z_j, m^z_{j+1}}.
\end{align}
Using these transformations to enact $\mathcal{T}$ on $\gamma_{1,2}$, one finds that the contribution of each term in the string of $s_{m^z_j, m^z_{j+1}}$ signs cancels with the next, except at the very ends of the chains.  
One thus recovers Eq.~\eqref{TransformationRules} as claimed.

\end{document}